\documentclass[useAMS,fleqn,usenatbib]{mnras}

\usepackage{newtxtext,newtxmath}

\usepackage[T1]{fontenc}
\usepackage{ae,aecompl}

\usepackage{graphicx}	
\usepackage{amsmath}	
\usepackage{amssymb}	
\usepackage[usenames]{color}
\def\mpc{~\rm{Mpc}}

\title[Filaments in VIPERS]{Filaments in VIPERS: galaxy quenching in the infalling regions of groups}

\author[Salerno, Mart\'inez \& Muriel]{
Ju\'an Manuel Salerno,$^{1}$\thanks{jsalerno@oac.unc.edu.ar}
H\'ector J. Mart\'inez,$^{1,2}$
and Hern\'an Muriel$^{1,2}$
\\
$^{1}$Instituto de Astronom\'{\i}a Te\'orica y Experimental (IATE), CONICET - UNC, Laprida 854, X5000BGR, C\'ordoba, Argentina\\
$^{2}$Observatorio Astron\'omico, Universidad Nacional de C\'ordoba, Laprida 854, X5000BGR, C\'ordoba, Argentina\\
}

\date{Accepted 2018 December 09. Received 2018 December 07; in original form 2018  September 29}

\pubyear{2018}

\hypersetup{draft} 
\begin{document}
\label{firstpage}
\pagerange{\pageref{firstpage}--\pageref{lastpage}}
\maketitle
 
\begin{abstract}
We study the quenching of galaxies in different environments and its evolution in the redshift range 
$0.43\le z \le 0.89$.  
For this purpose, we identify galaxies inhabiting in filaments, the isotropic infall region of groups, 
the field, and groups in the VIMOS Public Extragalactic Redshift Survey (VIPERS). 
We classify galaxies as quenched (passive), through their $NUV-r$ vs. $r-K$ colours. 
We study the fraction of quenched galaxies ($F_r$) as a function of stellar mass and environment
at two redshift intervals. Our results confirm that stellar mass is the dominant factor 
determining galaxy quenching over the full redshift range explored.
We find compelling evidence of evolution in the quenching of intermediate mass galaxies 
$(9.3 \leq \log(M_{\star}/M_{\odot}) \leq 10.5)$ for all environments.  For this mass range, $F_r$ is largest for galaxies in groups, and smallest 
for galaxies in the field, while galaxies in filaments and in the isotropic infall regions appear to have intermediate values with the exception of the high redshift bin, where the latter show similar fraction of quenched galaxies as in the field. Galaxies in filaments are systematically more quenched than their counterparts infalling from other directions, in agreement to similar results found at low redshift.  The least  massive galaxies in our samples, do not show evidence of internal or environmental quenching.
 \end{abstract}




\section{Introduction}

The large-scale structure of the Universe in the $\Lambda$CDM model is characterised by the 
anisotropic structure of the matter distribution, where we observe walls, filaments and their nodes. 
In the nodes galaxy clusters and groups are observed \citep{bond96}. Filaments trace the cosmic web 
connecting the nodes framing walls separated by large voids \citep{aragon10}.
After galaxy clusters and groups, filaments constitute the densest environment and
can account for up to $\sim 40\%$ of the Universe's mass \citep{aragon10}.

Several algorithms have been developed to identify filaments in the galaxy distribution. Some of them 
identify filaments as ridges in the density field (e.g. \citealt{Novikov06,aragon07,Sousbie11,Cautun13}).
Other algorithms use a minimal spanning tree \citep{Barrow85} method instead (e.g. \citealt{Colberg07,Alpaslan14}). 
Filaments are also detected as overdensities in the galaxy distribution between pairs of 
clusters \citep{Pimbblet05}, or groups \citep{Martinez16}.For a detailed comparison of the different web finders, see \citet{Cautun:2014} and references therein. Using numerical simulations, these authors also studied the evolution of the cosmic web and found that while most filaments have been in place since $z\sim ~0.5$, they have been evolving rapidly at higher redshift.

It is well known that galaxy properties strongly depend on stellar mass and environment 
(e.g., \citealt{dressler80,gomez03,martinez08}). 
Galaxy properties in clusters and groups have been studied thoroughly at different redshifts 
(e.g. \citealt{martinez06,martinez08,wilman09,kovac10,pressoto12,Cucciati17,coenda18}), 
however, the study of galaxies inhabiting the region between two groups/cluster 
(filament region) have not been as widely studied. 

The Galaxy And Mass Assembly (\citealt{Driver:2009}, GAMA) survey of the nearby Universe has been extensively used to study the properties of galaxies on large-scale environments.  \citet{Eardley:2015} investigated the galaxy luminosity function and found that the luminosity function's characteristic magnitude brightens 0.5 mag from voids to knots (groups/clusters). On the other hand, \citet{Alpaslan:2016} found that the stellar mass is the primary factor to predict the star formation rate (SFR), being the large-scale environment a second order parameter (see also \citealt{Alpaslan:2015}). Also using GAMA, \citet{Kraljic:2018} investigated the role of the cosmic web in shaping galaxy properties. They found that the fraction of red galaxies increases when approaching nodes and filaments. Using the Sloan Digital sky Survey (SDSS), \citet{Chen:2017} studied the effect of filaments on galaxy properties. They found that a red galaxy or a high-mass galaxy tends to reside closer to filaments than a blue or low-mass galaxy.

At intermediate redshifts, \citet{zhang13} studied the colour of galaxies located in between pairs of clusters in the range $0.12 \leq z \leq 0.4$ and detect an evolution in the blue fraction of filament galaxies that is not observed in clusters. They also found that richer clusters are connected to richer filaments.
\cite{Martinez16} studied the effects of environment upon galaxies infalling into groups in the SDSS DR7
\citep{Abazajian09} by distinguishing whether they are in filaments (filament galaxies, hereafter FG)
or falling from other directions (isotropic infalling galaxies, hereafter IG). 
They show evidence of galaxy preprocessing in the infall region of groups, and a 
distinct action of the filament environment upon galaxies. FG and IG galaxies differ 
from field galaxies and galaxies in groups in terms of their luminosity function and
their specific star formation rate. The authors found that, while the luminosity function of FG and 
IG galaxies are similar, they are intermediate between the luminosity function of field galaxies and 
that of galaxies in groups. They also found systematic lower values of specific star 
formation rate of FG compared to IG, thus providing evidence that galaxies infalling alongside 
filaments have experienced stronger environmental effects than galaxies infalling from other directions.

At higher redshifts, \cite{Darvish14} studied a sample of $H_\alpha$ emitting star forming galaxies at 
$z$ = $0.845$ in the COSMOS field \citep{Scoville07b}. They show an enhancement of star formation activity 
in filaments compared to denser regions (cluster) which they explain in terms of galaxy -- galaxy 
interactions.
\cite{Laigle17}, using a sample of galaxies from the COSMOS field, with photometric redshifts 
$0.5 \leq z_{\rm phot} \leq 0.9$, found that passive galaxies are more confined towards the core of 
the filament, in contrast to star-forming ones. \cite{Malavasi17} identified filaments in the final data 
release of  the VIMOS Public Extragalactic Redshift Survey (VIPERS, \citealt{Guzzo14}),
finding a significant segregation in the sense of the most massive galaxies 
likely to be closer to filaments. \citet{Davidzon:2016} studied how the environment 
shapes the stellar mass function using VIPERS. They focused on the large scale density and found 
that  the stellar mass function  evolves in high densities regions, while it stays nearly constant in low-density environments.

\cite{just15}, studied galaxies in the infall regions of 21 cluster at intermediate redshift 
($0.4 \leq z \leq 0.8$) from the ESO Distant Cluster Survey \citep{White05}.
They found that the fraction of red galaxies in the infall region is intermediate to that in clusters and 
in the field, suggesting a preprocessing of the properties related to the 
star formation outside the virial radius. 

In this paper, we search for filaments defined between pairs of groups at redshifts 
$0.43\leq z\leq 0.9$ in the final data release of VIPERS (PDR-2, \citealt{Scodeggio:2018}) in order to
study the effects of environment in quenching star formation in galaxies 
infalling into groups. Our goal is to statistically compare galaxies in the outskirts of 
groups distinguishing between those that are in regions where filaments are likely to be present, 
and those that are not. It is out of the scope of this paper to construct a complete catalogue of 
filaments, or to assess univocally if a particular galaxy belongs to a filament or not.

This article is organised as follows: we describe the galaxy data in Sect. \ref{sect:data}; 
Sect. \ref{sect:filam} deals with the identification of groups and filaments;
in Sect. \ref{sect:results} we compare the fraction of passive galaxies in groups, field, filaments
and the isotropic infalling region, and its redshift evolution. We discuss the implications of our
results in this section as well. Finally, we summarise our main results in Sect. \ref{sect:conclu}.
Throughout the paper we assume a flat cosmology with density parametres 
$\Omega_{\rm m} = 0.30$, $\Omega_{\Lambda} = 0.70$, and a Hubble's constant $H_0 = 70$ km s$^{-1}$ 
Mpc$^{-1}$. Unless otherwise specified, all magnitudes and colours are in the Vega system.

\section{Data}
\label{sect:data}

To identify groups of galaxies and filaments, we select a sample of galaxies from VIPERS PDR-2.
This final data release contains spectra for $\sim 90,000$ galaxies.
VIPERS is a deep spectroscopic galaxy survey on the 
23.5 deg$^2$ over the W1 and W4 fields of the \textquotesingle T0005\textquotesingle ~release of the 
Canada-France-Hawaii Telescope Legacy Survey Wide and 
completed with subsequent \textquotesingle T0007\textquotesingle. Covers the redshift range 
$0.45 \leq z \leq 1.4$ resulting in a volume of $1.46\times 10^8$ Mpc$^3$.

The target catalogue consisted basically of all objects within the magnitude range $i_{AB}\leq 22.5$ 
with a $(r-i)$ vs $(u-g)$ colour pre-selection to remove galaxies at $z < 0.5$. 
Spectra were collected with the VIMOS multi-object spectrograph \citep{lefevre03} at moderate 
resolution ($R = 220$) using the LR Red grism, leading to a radial velocity error of 
$\sigma_v = 175(1 + z_{\rm spec})$ km s$^{-1}$. We refer the reader to \citet{Garilli14} for a complete
description of the survey data, and to \citet{Guzzo14}  for specifications regarding target selection 
and survey design.

We restrict our sample to galaxies with 
high-quality redshift measurement, i.e., galaxies that have flag $\geq 2$ according to \cite{Scodeggio:2018}. This cut-off results in a sample with a redshift 
confirmation rate of $96.1 \%$. In total,  W1 and W4 fields of our sample comprises $72,369$ galaxies.

In addition to the VIPERS spectroscopic sample, we make use of the T0005 release of the CFHTLS  
Survey\footnote{http://www.cfht.hawaii.edu/Science/CFHTLS/} in the $u', g',r',i'$, and, $z'$ bands, matched 
with the final CFHTLS release (T0007) in the $u, g, r, i$, and $z$ bands. It also includes
$FUV$ and $NUV$ magnitudes from GALEX \citep{Martin05}, and $Ks$-band photometry from WIRCam \citep{Puget:2004}. 
For all survey details and the full photometry see VIPERS Multi-Lambda Survey \citep{Moutard:2016}.


We computed absolute magnitudes and stellar masses using the code {\it Hyperzmass}, a modified
version of {\it Hyperz} \citep{Bolzonella00}, which uses a spectral energy distribution (SED) fitting technique.
We used a library of SEDs derived from the stellar population synthesis model (SSPs) by \citet{Bruzual03}. 
Our library consists in several exponential decaying SFR models, with characteristic timescales:
0.1, 0.3, 0.6, 1, 2, 3, 5, 10, 15, and 30 Gyr; and two different metallicities: 
$Z=Z_{\odot}$ and $Z =0.2 Z_{\odot}$. We assume a Chabrier initial mass function \citep{Chabrier03} and generate these model 
templates by means of the CSP routine of GALAXEV \citep{Bruzual03}. 
The dust content of the galaxies was modelled using the Calzetti's law \citep{Calzetti:2000} option in 
{\it Hyperzmass}, allowing the extinction in the $V-$band to range from 0 to 3 mag. 


To study how environment and stellar mass quench star formation in galaxy we rely on the  
NUV$-r$ vs. $r-K$ colour-colour diagram, $NUVrK$, to classify galaxies according to their
stellar populations. \citet{Arnouts13} showed that the $NUVrK$ diagram is a good alternative 
to assess the total SFR 
of star-forming galaxies without involving complex modeling such as SED fitting techniques.
In particular, we use the criterion by \citet{fritz14} (their Fig. 7) in the $NUVrK$: 
if a galaxy's colours are such that $NUV-r > 1.25 \times (r-K) - 0.9$,
then it is considered to be a red passive one.

c\section{Filament identification and environment classification}
\label{sect:filam}

\subsection{Galaxy groups}

Our filament identification algorithm searches for galaxy overdensities in the regions between
groups of galaxies. Thus, our starting point to find filaments in VIPERS is identifying groups
of galaxies first. For group identification we use a friends-of-friends (FOF, \citealt{Huchra82})
that links galaxies into groups.
Particularly we use an algorithm likely-FOF following \citet{Eke04}. 
This algorithm scales both, the perpendicular (in the plane
of the sky), and the parallel (line-of-sight) linking lengths ($l_{\perp} $, and
$l_{\parallel}$, respectively) as a function of the observed mean density of galaxies, as $n(z)^{-1/3}$.
The FOF has three free parameters that are usually optimized with the aid of a mock galaxy catalogue: the linking 
length $b$, the maximum perpendicular linking length $L_{\mbox{max}}$, which is introduced to avoid unphysically 
large values for $l_{\perp}$, and the ratio between the linking length along and perpendicular
to the line of sight $R$ (see \cite{Eke04} for more details).
\cite{Knobel12} identify galaxy groups with an adapted FOF as in \cite{Eke04} over the 20k zCOSMOS-bright 
redshift survey \citep{Lilly07}. This survey covers the redshift range $0.1 \leq z \leq 1.0$.
In this paper we adopt the parameters used by \cite{Knobel12}, because of the similarity
of the redshifts involved.
	
We compute physical parameters of the groups: the velocity dispersion and the virial radius.
Since $99\%$ of our galaxies groups have less than 15 members, we estimate the velocity dispersion with the 
gapper estimator \citep{Beers90}.
We identify a total of 1013 galaxy groups. Projected virial radii are computed as twice the harmonic mean of the 
projected distances between group members.
Our sample of groups have median virial radius of 1.0 Mpc, and median velocity dispersion of 
350 km s$^{-1}$.

\subsection{Filament identification}

\begin{figure}
	\includegraphics[scale=0.55]{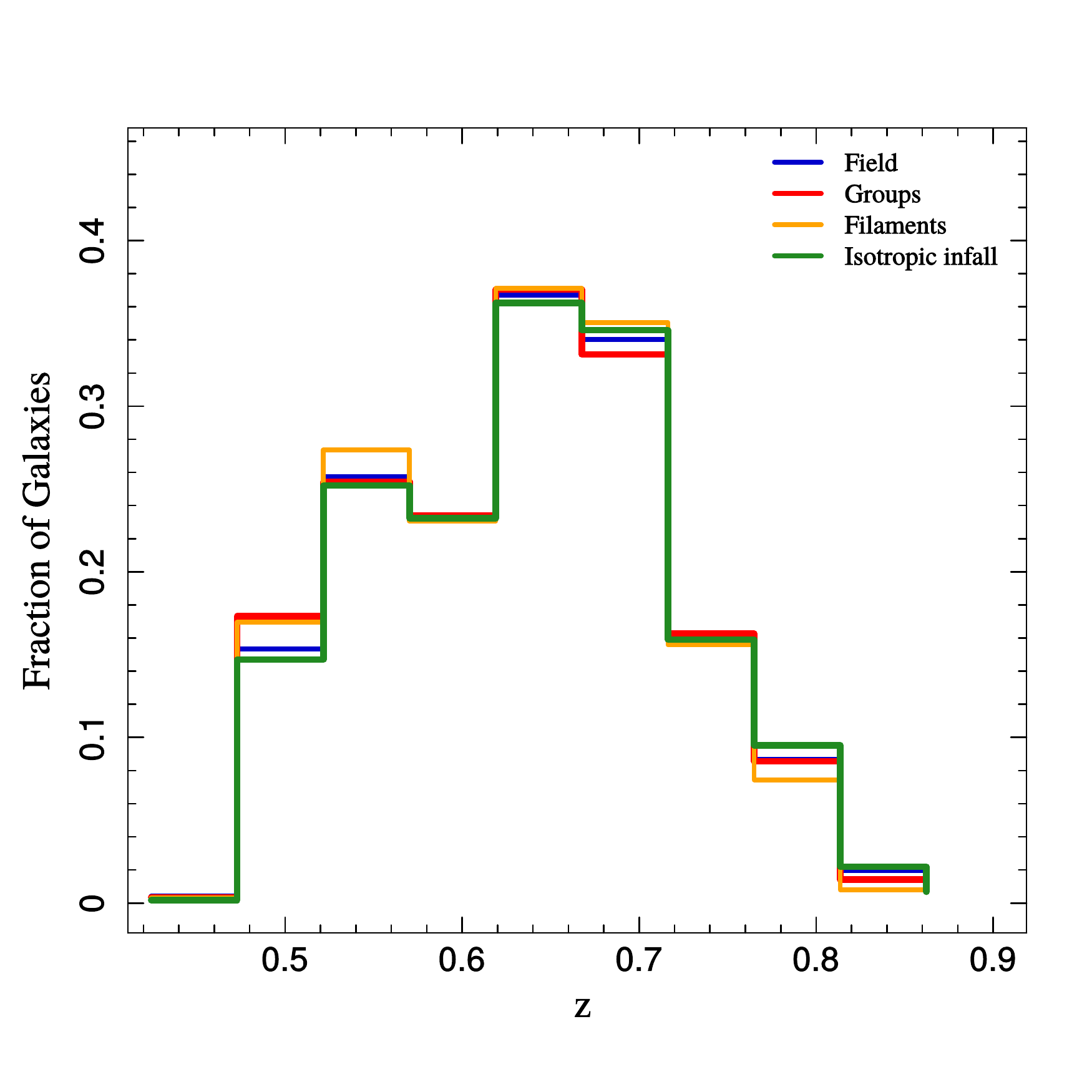}
    \caption{Normalised redshift distributions of the different samples of galaxies.}
    \label{fig1}
\end{figure}

We identify filamentary structures that extend between groups of galaxies following \citet{Martinez16}.
In what follows, unless otherwise specified, we make all computations using comoving distances in redshift 
space.

Firstly, we select candidates to be 
filaments nodes, which are pairs of groups ($i,j$) meeting the following criteria: 
1) the projected distance (computed at the mean redshift) between them, $\Delta \pi_{ij}$, is smaller 
than a chosen value $\Delta \pi_{\text{max}}$, yet larger than the sum of their projected 
virial radii, $r^{(i)}_{\rm pv} + r^{(j)}_{\rm pv} < \Delta \pi_{ij} \leq \Delta \pi_{\rm max}$; 
2) the line-of-sight distance between them, 
$\Delta \sigma_{ij}$, is smaller than certain $\Delta \sigma_{\rm max}$. 
The first condition ensures that the nodes are clearly separated structures in projection.
In this work we use $\Delta \pi_{\rm max}= 15 \mpc$ and $\Delta \sigma_{\rm max}=15 \mpc$.

Once groups $i$ and $j$ satisfy conditions 1) and 2) above, we use an overdensity criterion to decide whether 
there is a filament connecting them. We compute the galaxy overdensity in a rectangular cuboid with one 
of its axes aligned along the line-of-sight ($z-$axis), and whose base is in the plane of the sky
with one of its sides parallel to the projected separation between the groups ($x-$axis), 
and the other perpendicular ($y-$axis). 
From the mean redshift of the pair, this rectangular cuboid extends $\pm 15 \mpc$ alongside its 
$z-$axis.
The base is positioned in order to cover the projected region between the two groups, while not
overlapping with them (Fig. 1 of \citealt{Martinez16}). 
Its dimensions are: $\Delta \pi_{ij}-r^{(i)}_{\rm pv}-r^{(j)}_{\rm pv}$
in the $x-$axis, and $\pm 1.5 \mpc$ in the $y-$axis. The overdensity, $\delta =n/n_r-1$, 
is computed by counting VIPERS galaxies in 
this region ($n$), and points from a random catalogue that has the same angular mask 
and redshift distribution of VIPERS ($n_r$). Our random catalogue is $100$ times denser than
VIPERS, thus we rescale $n_r$ accordingly. If $\delta>1$, we consider the pair is
linked by a filament \citep{Martinez16}.
Out of 718 candidate pairs, we find 656 filaments.

\subsection{Defining environments}

\begin{figure*}
	\includegraphics[scale=1]{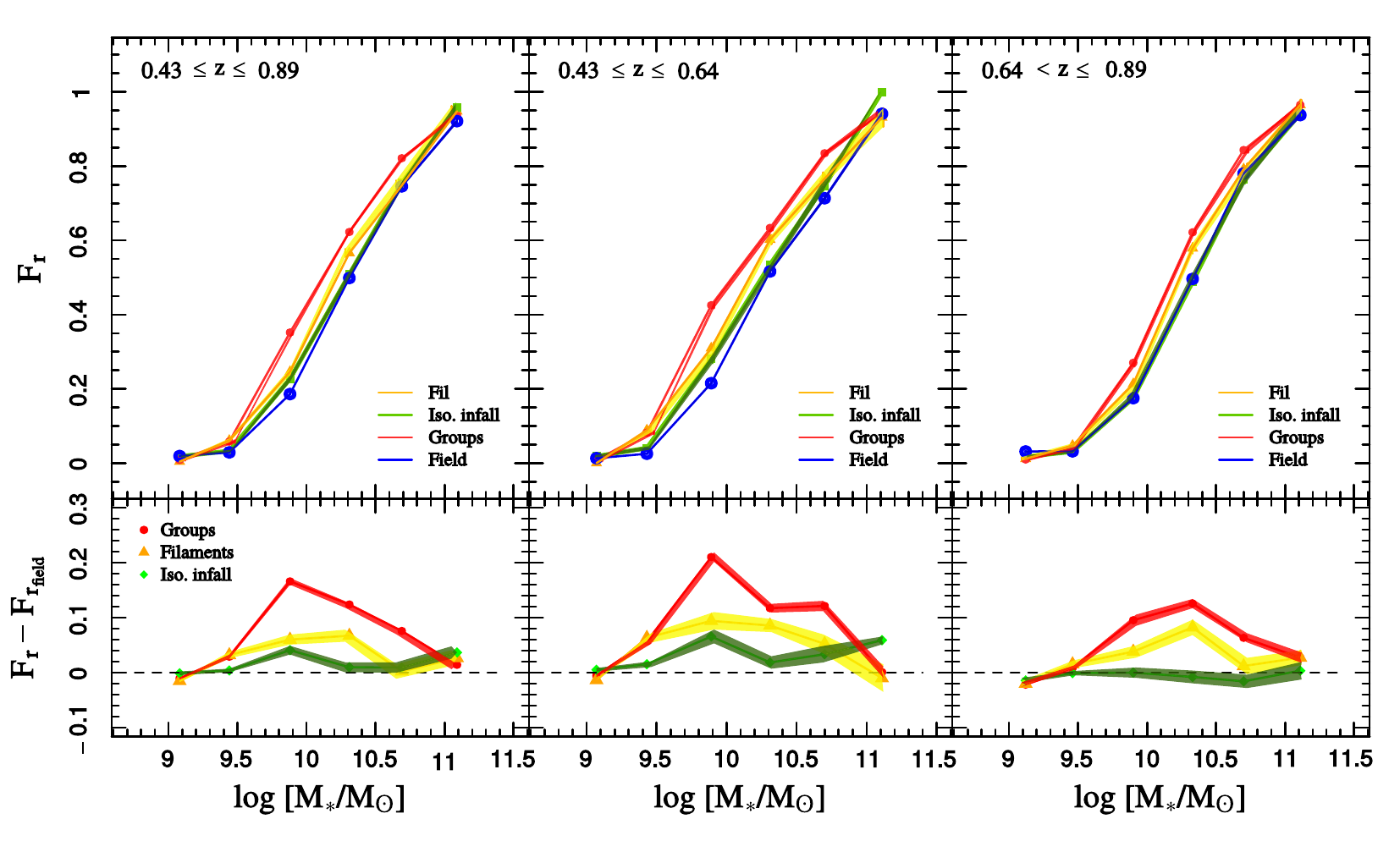}
    \caption{Red galaxies fraction as a function of stellar mass for our samples of galaxies.
    {\em Left panel} corresponds to the entire redshift range, {\em central panel} to the first redshift bin and the 
    {\em right panel} to the high redshift bin. Colours denote environment: {\em blue} corresponds to
    field galaxies, {\em green} to isotropic infalling galaxies, {\em dark yellow} to galaxies in filaments,
    and {\em red} to galaxies in groups. Error-bars were computed with the bootstrap resampling technique.}
    \label{fig2}
\end{figure*}

The identification of filaments in VIPERS of the previous subsection determines three of
our four subsamples which we use in our analyses below: galaxies in the groups that define
the filaments (GG), galaxies in the filaments (FG), and galaxies in the infall regions of those groups
that are not located in filaments (IG).

The sample of galaxies in groups (GG) is made of all VIPERS galaxies that were identified as members
of the groups that are linked by filaments. This sample comprises 7407 galaxies.

The sample of galaxies in filaments (FG) includes all galaxies that were found to be in the 
filament regions identified in the previous subsection. Every FG is associated to its closest
node. This means that each node in the filament defined by groups $i$ and $j$ 
contributes to the FG sample with galaxies that are as far as $\Delta \pi_{ij}/2$ in projection.
Our sample of FG consists of 2311 galaxies.

We use the projected distance $\Delta \pi_{ij}/2$ to define the isotropic infall region around each 
group: a cylinder centred in the group and aligned along the line of sight direction, whose 
radius is $\Delta \pi_{ij} /2$, and whose height is $2 \,\Delta \sigma_{\text{max}}$  (see  
Fig. 1 of \citealt{Martinez16}). The isotropic infall sample amounts to 3931 galaxies.

Nodes can be one of the extreme point of more than one filament, thus they may contribute to both,
FG and IG samples, more than once and up to different projected distances. We check the samples
in order not to have repeated galaxies.
By construction, the GG, FG and IG samples have similar redshift distributions, as can be seen
in Fig. \ref{fig1}. 

We also construct a sample of field galaxies (CG, C for control), by means of a Monte Carlo
algorithm that randomly selects VIPERS galaxies in order to create a sample that has a similar redshift
distribution as the previous three samples. We take special care of avoiding in the process all 
galaxies in groups, in filaments, and in the isotropic infall regions. Our CG sample
comprises 30,299 galaxies. The normalised galaxy redshift distributions for the four environments probed are shown in Fig. 1.

\begin{figure*}
	\includegraphics[scale=0.80]{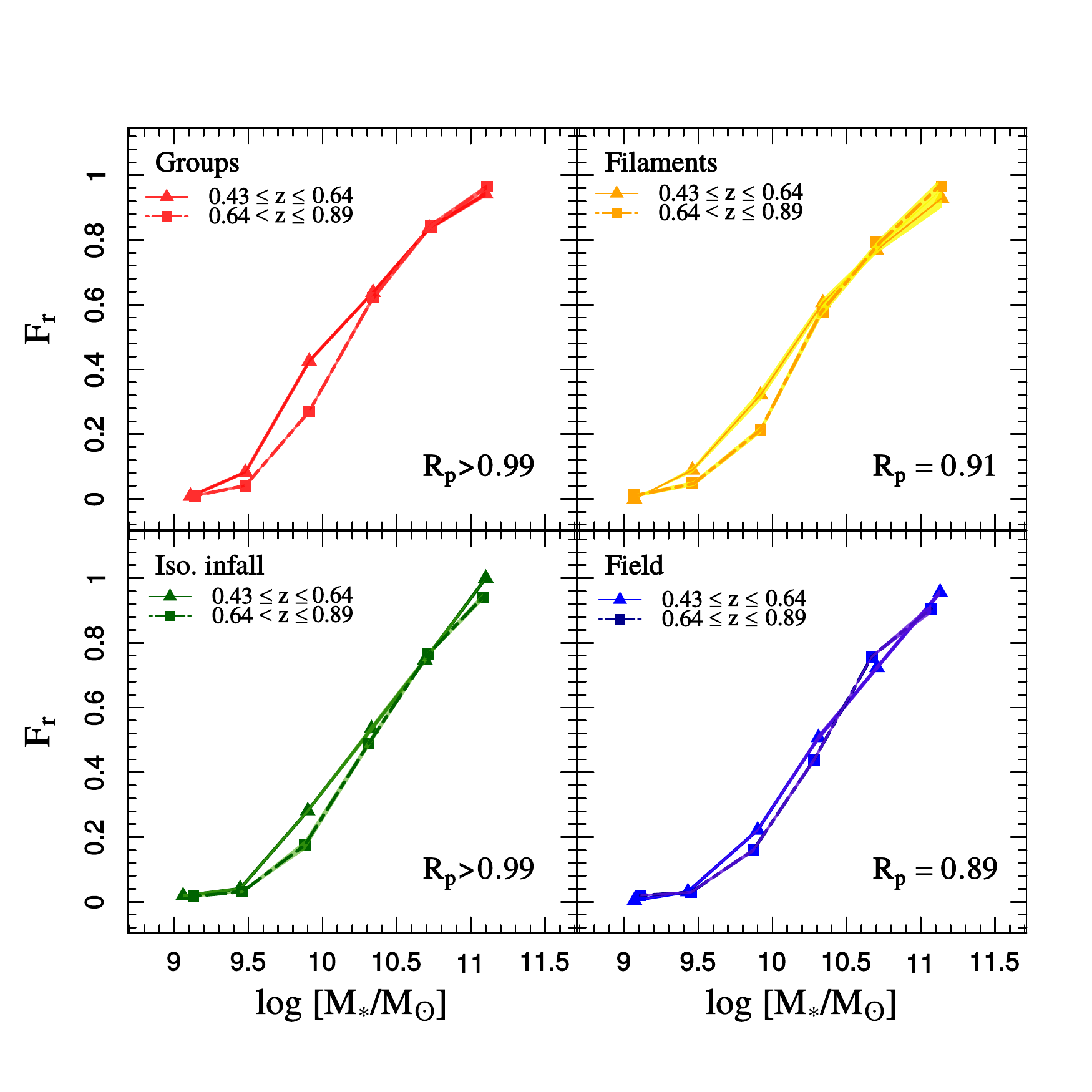}
    \caption{Passive galaxy fraction evolution. {\em Top left panel} shows galaxies in groups, 
    {\em Top rigth panel:} galaxies in filaments, {\em bottom left} and {\em right panel} 
    galaxies in the isotropic infall and field region,
    respectively. Solid lines correspond to the low-redshift bin, dashed lines to the high-redshift bin.
    Also, we show the results of applying the test of \citet{Muriel:2014} (see text) to check whether the
    redshift evolution is statistically significant in each case.}
    \label{fig3}
\end{figure*}

\section{RESULTS}
\label{sect:results}

\begin{table}
\begin{center}
\begin{tabular}{cccc}
\hline
 & \multicolumn{3}{c}{Rejection probability} \\
Redshift range          &  FG vs. IG     &  FG vs. CG     &   IG vs. CG \\
\hline 
 $0.43 \leq z \leq 0.64$   &  0.92     & >0.99  & >0.99   \\ 
 $0.64 < z \leq 0.89$      &  0.96  & >0.99  & 0.78   \\
\hline  
\end{tabular}
\end{center}\caption{Results of the test that computes the cumulative differences between the fractions of red galaxies of two different samples along the stellar mass domain (see Fig. \ref{fig2}). The quoted numbers are the rejection probability of the null hypothesis of two samples as drawn from the same underlying distribution.}
\label{table1}
\end{table}

We split our samples of galaxies into two redshift bins, defined by the median redshift 
$z_{\rm med}=0.64$, and then compute for each environment and redshift bin, the fraction
of red galaxies as a function of stellar mass.
These fractions can be seen in the upper panels of Fig. \ref{fig2}, where the left panel 
compares the four environments in the entire redshift range, the central does it at the
first redshift bin ($0.43 \leq z \leq 0.64$), and the right panel corresponds to the second 
redshift bin ($0.64 < z \leq 0.89$).
Lower panels in Fig. \ref{fig2} show the differences in the fraction of quenched galaxies
relative to the field values for a better visualization of the effects of the three densest
environments we study. It should be remembered that, by construction, our samples of IG and FG are  
contaminated by field galaxies. 

It is clear from Fig. \ref{fig2} that stellar mass is the dominant factor behind
galaxy quenching, as it has been extensively discussed in the literature (see for instance \citealt{Peng:2010}). 
Independently of redshift and environment, the fraction of red (quenched)
galaxies is a strong function of stellar mass. 

In general, at the low mass end ($\log(M_{\star}/M_{\odot})\sim 9.1$), all environments appear to be 
ineffective to quench galaxies. On the contrary, at the highest mass bin 
($\log(M_{\star}/M_{\odot})\sim 11.1$) all environment have a similar fraction of quenched galaxies,
which indicates that galaxies this massive are likely to be quenched almost exclusively by internal
processes. For masses in between those extremes, the differential effects of environment are evident
at both redshift bins. As an overall trend, at fixed mass, GG have the highest fraction of 
quenched galaxies, followed by FG, and the lowest fraction is seen for field galaxies. 
The fraction of quenched galaxies in IG and CG are indistinguishable at our highest redshift bin.
On the other hand, at our lowest redshift bin, the fraction of quenched IG and CG differ.
With the former having an intermediate behavior between FG and CG.

Several authors (e.g. \citealt{Sobral:2011,Muzzin:2012,Darvish:2016}) have presented evidence
that the effects of environment upon galaxy quenching became noticeable by $z\sim 1$. 
The upper right panel of Fig. \ref{fig2} shows evidence of environmental quenching by filaments
already present at $0.64 < z \leq 0.89$. The effects of the isotropic infall region became
significant later on cosmic time, they are only present at our lowest redshift bin. This may be 
indicating that this environment was not dense enough at the highest redshift bin to produce any 
noticeable effect upon galaxies. This distinction between FG and IG, was reported
at low redshift ($z\sim 0.1$) by \citet{Martinez16}. 

We test whether differences seen in this figure are statistically significant by using a similar
test to that of \citet{Muriel:2014}. When comparing two trends in Fig. \ref{fig2}, this test computes the 
cumulative differences between the fractions of red galaxies in the two samples along the stellar mass 
domain, and then checks whether the resulting quantity is consistent with the null hypothesis of 
the two samples drawn from the same underlying population. Results of the test for FG, IG and CG 
(GG are clearly different from these three) are quoted in Tab. \ref{table1} in terms of the rejection 
probability of null hypothesis ($R_{\rm p}$). 
This test confirms that the null hypothesis is overruled in all cases, with the exception of IG and CG at
the highest redshift bin. The fact that the fraction of quenched IG and CG do not differ 
may be indicating that the isotropic infalling region around groups were an environment not different
from the field at these redshift. On the other hand, filaments were already a distinctive environment
at these times.

For a better visualization of redshift evolutions, all trends in Fig. \ref{fig2} are re-arranged 
in Fig. \ref{fig3}, where we compare each environment
with itself at the two redshift bins.
In all four cases, there is clear evidence of an increment in the fraction of quenched galaxies 
with decreasing redshift. This evolution is seen for intermediate mass galaxies,
$9.1 \lesssim \log(M_{\star}/M_{\odot}) \lesssim 10.4$, and is practically 
absent outside this range. For higher masses, all galaxies in 
all environments were already quenched at the high redshift bin;
at the lowest mass bins there are almost no quenched galaxies at any of the two redshift bins.
The latter means that galaxies with stellar masses $\log(M_{\star}/M_{\odot})\lesssim 9.1$ were  
difficult to quench at these times of the history of the Universe in the environments under consideration.

We quote in Fig. \ref{fig3} the results of applying the test described above, 
to check whether the evolution of the fraction of red galaxies in each environment are statistically 
significant. In terms of of the rejection probability of null hypothesis, the fraction of red galaxies 
in GG at the two redshift bins are distinguible ($R_{\rm p}> 0.99$), a similar result is found for 
the IG ($R_{\rm p} > 0.99$), which supports the case for evolution in these two environments. 
The test is not that conclusive for FG ($R_{\rm p}=0.91$) nor for CG ($R_{\rm p}>0.89$).

There is another interesting feature in Fig. \ref{fig3}: the largest evolution in the fraction of 
quenched galaxies occurs at $\log(M_{\star}/M_{\odot})\sim 9.9$ irrespective of environment, with an
almost null evolution for masses $\log(M_{\star}/M_{\odot})>10.3$. This evolution is stronger as we move
from the field to higher density environments. This suggests that in the redshift range we probe,
the galaxies that are the most likely to be quenched are those with mass 
$\log(M_{\star}/M_{\odot})\sim 9.9$. This should be due to a combination of internal and external 
factors.
External factors are clearly present since the strength of the evolution depends on environment. 
On the other hand, the fact that, independently of environment, the largest change occurs at the same 
mass range, may be pointing out to these galaxies as the best prepared to be quenched, and this could
be preferentially attributed to internal processes (mass quenching).

Another conclusion worth mentioning that can be deduced from our results is that galaxies that fell 
into groups at least as far back in time as $z\sim 0.9$, were likely to be pre-processed by the 
environment surrounding the systems. The likelihood of being pre-processed was higher if 
these galaxies happened to fell along filaments. 

\section{Summary}
\label{sect:conclu}

In this work we study the effects of environment on galaxy quenching at $0.43\le z\le 0.89$
using data from VIPERS. Firstly, we identify groups of galaxies and filamentary structures 
stretching between them. We study the fraction of passive galaxies (defined by their 
$NUV-r$ vs. $r-K$ colours) as a function of stellar mass in 4 different environments: field, 
filaments, the isotropic infalling region of groups, and groups. We also studied, for each 
environment, the evolution of the fraction of quenched galaxies within the redshift range 
covered by VIPERS.

The effects of environment are 
significant for intermediate mass galaxies: the fraction of quenched galaxies
is lowest in the field, higher in the isotropic infall region, even higher in filaments, and 
finally the highest values are reached in groups. At the two extremes in stellar mass, we 
found opposite situations: on the one hand, massive galaxies 
do not appear to require the action of environment to be quenched; on the other hand, the least 
massive galaxies in our samples, are hardly affected by the environment at all.

We find evidence of environmental quenching in galaxies in filaments at least from $z\sim 0.9$,
and in the isotropic infall region of groups from $z\sim 0.6$ onwards. 
There is evolution in the fraction of intermediate mass galaxies in the redshift range under study.
This evolution is stronger in groups, followed by filaments, the isotropic infall region,
and the mildest (if any) evolution is seen in the field.

\section*{Acknowledgements}

This paper has been partially supported with grants from Consejo Nacional de 
Investigaciones Cient\'ificas y T\'ecnicas (PIP 11220130100365CO) Argentina, 
and Secretar\'ia de Ciencia y Tecnolog\'ia, Universidad Nacional de C\'ordoba, Argentina.
This research makes use of the VIPERS-MLS database, operated at CeSAM/LAM, Marseille, France. This work is based in part on observations obtained with WIRCam, a joint project of CFHT, Taiwan, Korea, Canada and France. The CFHT is operated by the National Research Council (NRC) of Canada, the Institut National des Science de l'Univers of the Centre National de la Recherche Scientifique (CNRS) of France, and the University of Hawaii. This work is based in part on observations made with the Galaxy Evolution Explorer (GALEX). GALEX is a NASA Small Explorer, whose mission was developed in cooperation with the Centre National d'Etudes Spatiales (CNES) of France and the Korean Ministry of Science and Technology. GALEX is operated for NASA by the California Institute of Technology under NASA contract NAS5-98034. This work is based in part on data products produced at TERAPIX available at the Canadian Astronomy Data Centre as part of the Canada-France-Hawaii Telescope Legacy Survey, a collaborative project of NRC and CNRS. The TERAPIX team has performed the reduction of all the WIRCAM images and the preparation of the catalogues matched with the T0007 CFHTLS data release.




\bibliographystyle{mnras}
\bibliography{LEGACY/refs} 


\appendix


\bsp	
\label{lastpage}
	\end{document}